# The electron gas in a magnetic field: the equation of state

Vladan ^elebonovi}

Institute of Physics, Pregrevica 118,11080 Zemun- Beograd, Yugoslavia

celebonovic@exp.phy.bg.ac.yu
vcelebonovic@sezampro.yu

Abstract: Continuing previous work and using a simple change of variables, an analytical equation of state for a degenerate non-relativistic Fermi gas in a magnetic field is proposed.

## Introduction

This paper is a continuation of previous work (^elebonovi},1998a) on the equation of state (EOS) of a degenerate non-relativistic Fermi gas. In that paper, an analytical expresion for the Fermi-Dirac (FD) integrals of arbitrary order was proposed and applied in obtaining the EOS. The purpose of the present paper is to propose and analogous expression, but applicable to the case of a Fermi gas in a magnetic field. The nature of the source of the field is irrelevant, but the problem has considerable physical signficance, because magnetic fields occur in a wide variety of physical settings.

## Calculations

The FD integrals have the following form

$$F_n(\eta) = \int_0^\infty \frac{f(\varepsilon)d\varepsilon}{1+\exp[\beta(\varepsilon-\mu)]} \qquad (1)$$

where $f(\varepsilon) = \varepsilon^l, l \in R$, $\mu$ is the chemical potential, $\beta$ the inverse temperature and $\eta = \beta\mu$. The EOS of a degenerate Fermi gas is ( for example, Cloutman,1989 )

$$n = \frac{4\pi}{h^3}(2m_e k_B T)^{3/2} F_{1/2}(\eta) \qquad (2)$$

where all the symbols have their standard meanings.

It has recently been shown ( ^elebonovi},1998a) that the FD integrals of arbitrary order can be expressed as

$$F_n(\eta) = \int_0^m f(e)de + T\sum_{n=0}^{\infty} \frac{f^{(n)}(m)}{n!}[1-(-1)^n]T^n(1-2^{-n})\Gamma(1+n)V(1+n) \quad (3)$$

The symbols $\Gamma$ and $V$ denote the gamma and zeta functions.

How does this expression change when an external magnetic field is applied to he system? It has been proposed (Landau and Lifchitz,1976) that the influence of such a field can be described by the following change of variables:

$$m \to m \mp m_B H = m(1 \mp a) \quad (4)$$

where $m_B = \frac{|e|\hbar}{2m_e c}$ is the Bohr magneton, H is the magnetic field and $a = \frac{m_B H}{m}$.

The sign in (4) depends on the orientation of the electronic magnetic moments with respect to the field.

Inserting (4) into (3),and using the definition of $f(e)$, one gets the following form of a FD integral of arbitrary order in the presence of a magnetic field:

$$F_n(\eta) = \frac{[m(1\mp a)]^{l+1}}{l+1} + T\sum_{n=0}^{\infty} \frac{f^{(n)}[m(1\mp a)]}{n!}[1-(-1)^n]T^n(1-2^{-n})\Gamma(n+1)V(n+1) \quad (5)$$

Using the result for the chemical potential of the electron gas given in (^elebonovi},1998a) the parameter $a$ can be expressed as follows

$$a = \frac{m_B H}{m_0[1 - \frac{1}{12}\left(\frac{pT}{m_0}\right)^2 + \frac{1}{720}\left(\frac{pT}{m_0}\right)^4 - \frac{1}{162}\left(\frac{pT}{m_0}\right)^6 + \frac{1}{754}\left(\frac{pT}{m_0}\right)^8]} \quad (6)$$

In this equation $m_0 = A n_e^{2/3}$ and $A = (3p^2)^{2/3}\frac{\hbar^2}{2m_e}$. Inserting (6) into (5),choosing the sign + in eq.(4) one gets

$$F_{1/2}(\eta) \cong \frac{2}{3}[m_B H + An^{2/3} - \frac{(pT)^2}{12An^{2/3}} + \frac{(pT)^4}{720A^3 n^2} - \frac{(pT)^6}{162 A^5 n^{10/3}} + \frac{(pT)^8}{754 A^7 n^{14/3}}]^{3/2}(1+ << 3 >>)$$
(7)

In eq.(7) the symbol <<...>> denotes the number of omitted terms. Expanding eq.(7) in its full form, it follows that the FD integral of the order $\frac{1}{2}$ can be expressed as

$$F_{1/2}(h) \cong \frac{m_B H}{54\sqrt{1885}}\left[\frac{2442960 m_B H A^7 n^{14/3} + <<5>>}{A^7 n^{14/3}}\right] + <<23>> \tag{8}$$

Inserting eq.(8) in its full form into eq.(2), it follows that

$$n \cong 23.6954 m_B H (m_e T)^{3/2} \, Sqrt[(m_B H A^7 n^{14/3} + A^8 n^{16/3} - 0.822467 n^4 T^2 A^6 + 0.13529 A^4 n^{8/3} T^4 - 5.9345 A^2 n^{4/3} T^6 + 12.5843 T^8)/(A^7 n^{14/3})])/h^3 + <<23>> \tag{9}$$

This equation is too complex to be solvable analytically. However, the low temperature limit of the full form of this equation can be solved analytically. Developing eq.(9) in its full form in T up to and including terms of the order $T^2$, one arrives at the following implicit form of the EOS of the degenerate non-relativistic electron gas in a magnetic field

$$n \cong \frac{16p\sqrt{2}}{3h^3}(mn^{2/3}AT)^{3/2}\left(1 + \frac{m_B H}{An^{2/3}}\right)^{3/2} \tag{10}$$

Equation (10) can be solved analytically under the simplifying assumption that $x = \frac{m_B H}{An^{2/3}} \to 0$. Physically, this limits the region of applicability of the solution of eq.(10) to the weak field and/or high density regime. Developing eq.(10) into series in x up to and including second order terms, and after some algebraic manipulation, the following expression is obtained

$$(1 - BT^{3/2})y^2 = (Fy + G)T^{3/2} \tag{11}$$

where the following symbols were used

$$y = n^{2/3}; \; B = \frac{16p\sqrt{2}}{3}\left(\frac{m_e^{1/2}}{h}\right)^3 A^{3/2}; \; F = \frac{3B}{2}\left(\frac{m_B H}{A}\right); \; G = \frac{3B}{8}\left(\frac{m_B H}{A}\right)^2 \tag{12}$$

Solving eq.(11) within S. Wolfram's MATHEMATICA 2.2 software package, one gets

$$y = \frac{-(FT)^{3/2} \pm Sqrt[F^2T^3 - 4GT^{3/2}(-1+BT^{3/2})]}{2(-1+BT^{3/2})} \qquad (13)$$

which is the EOS of the non-relativistic degenerate Fermi gas in a magnetic field. This is the implicite form of the EOS. Using the change of variables defined in eq.(12) one could obtain the explicit form of the equation of state .

What about the applications of the EOS derived in this paper? Equation (13) was derived for the low-temperature regime, under the assumption that $\frac{m_B H}{An^{2/3}} \to 0$ .It follows from this assumption that (13) is applicable to the low temperature electron gas whose number density is limited by $n \gg \left(\frac{m_B H}{A}\right)^{3/2}$. In astrophysics, situations like this can be expected in accretion disks around compact objects, while in laboratory work they can arise in high pressure experiments performed under external magnetic fields ( like, for example, in studies of organic metals) .

55